\pdfoutput=1
\documentclass[12pt,letterpaper]{article}
\usepackage[margin=1in]{geometry}
\usepackage[T1]{fontenc}
\usepackage[utf8]{inputenc}
\usepackage[english]{babel}
\usepackage{amsmath,amssymb,amsthm,mathtools}
\usepackage{bm}
\usepackage{graphicx}
\usepackage{float}
\usepackage{booktabs}
\usepackage{caption}
\usepackage{subcaption}
\usepackage{setspace}
\usepackage[authoryear,round]{natbib}
\usepackage{hyperref}
\theoremstyle{remark}
\newtheorem{remark}{Remark}
\title{A Toolkit for the Study of Treatment-Effect Discontinuities%
  \thanks{This research received no specific grant from any funding agency.}}
\author{Alessandro Baldi Antognini \and Paolo Verme%
  \thanks{Corresponding author: Paolo Verme, Department of Statistical Sciences,
  University of Bologna, Via delle Belle Arti 41, 40126 Bologna, Italy.
  Email: \texttt{paolo.verme@unibo.it}.}\\
  {\small Department of Statistical Sciences, University of Bologna}}
\date{}
\begin{document}
\maketitle
\begin{singlespace}
\begin{abstract}
\noindent This paper provides a toolkit for the study of distributional treatment effects (DTEs) focused on treatment-effect discontinuities defined as points where marginal distributional effects change sign. Building on the Treatment Effects Curve (TEC, \citealp{Verme2010TEC}), the paper makes three contributions. First, we propose a methodological framework comprising a Horizontal Discontinuity Analysis (HDA) comparing groups in regions of opposite-signed effects using causal forests, and a Vertical Discontinuity Analysis (VDA) examining sign-switch points. Second, we adapt crossing-point asymptotics to locate where a TEC crosses zero and to test the non-tangentiality of its local slope with a bias-corrected Wald statistic. Third, we illustrate the full workflow on synthetic data and add a diagnostic application to Mexico's PROGRESA data. The paper shows how these contributions complement and expand existing instruments for DTE analyses.
\end{abstract}

\vspace*{0.5cm}
\noindent\textbf{JEL Classification:} C14, C21, C52, I38

\noindent\textbf{Keywords:} Distributional treatment effects; Stochastic dominance; Cumulative distribution functions; Quantile regression; Causal forest; Regression discontinuity
\end{singlespace}

\newpage

\section{Introduction}\label{introduction}
Over the past four decades, the literature on average treatment effects (ATEs) has progressively become the backbone of the impact evaluation literature with randomized controlled trials (RCTs) being considered the gold standard. More recently, research on distributional treatment effects (DTEs) has gained momentum in recognition of the fact that ATEs may well hide important heterogeneities in treatment effects irrespective of the data source. Thanks to advances in econometrics and machine learning, this new branch of research has leveraged these fields to develop specific models to study DTEs such as quantile regression models, causal random forests, and other nonparametric methods.

This paper follows in this latter tradition but shifts the focus to treatment effects discontinuities, particularly the analysis of sign discontinuities where marginal distributional effects change sign. We argue that this is one of the most informative aspects in DTE analysis and one that requires specific instruments. We propose to use the treatment effects curve (TEC, \citealp{Verme2010TEC}), the curve constructed by taking the difference between two cumulative distribution functions (CDFs) representing outcomes for treated and non-treated groups.

We make three distinct contributions to the DTEs literature. The first contribution is methodological. We develop two complementary analytical methods for studying treatment effect heterogeneity: 1) \emph{Horizontal Discontinuity Analysis} (HDA), a comparative analysis of groups located in regions where marginal distributional effects have opposite signs using causal forest (CF), and 2) \emph{Vertical Discontinuity Analysis} (VDA), an analysis focused on the points where distributional curves cross, based on a local discontinuity design. We discuss the identifying assumptions required for causal interpretation and highlight important caveats regarding the use of endogenously determined thresholds.

The second contribution is inferential. We adapt existing crossing-point asymptotics to the TEC setting and operationalize an inferential procedure for the two objects on which discontinuity analysis hinges: the location of the points where the TEC crosses zero, estimated from the un-smoothed empirical CDFs, and the local slope of the TEC at those points, estimated by bias-corrected kernel smoothing. A Monte Carlo study documents its finite-sample performance including detection rates for steep and flat crossings, false positive rates under a placebo design with no crossing, bias and RMSE of the crossing-point estimator, the coverage of the proposed confidence intervals, and the gains in slope recovery and location-interval calibration from bias-correcting the local slope.

The third contribution is empirical. We illustrate the full workflow using synthetic data provided by the World Bank for simulation purposes, demonstrating the practical implementation of TECs, HDA, and VDA in a setting where the crossing structure is known by construction. We then add a compact diagnostic application to PROGRESA, Mexico's conditional cash transfer program, using the public data of \citet{angelucci2009}. TECs allow researchers to focus on distributional sign changes that may matter for policy, while the inference procedure helps determine whether threshold-based HDA and VDA analyses are warranted. The PROGRESA diagnostic is especially useful because it shows how subtle these changes can be in real data. Average effects and even standard distributional summaries can suggest a simple story, while the TEC reveals weak, localized, and dynamically changing spillover patterns.

The paper is organized as follows. Section 2 reviews recent advances in the analysis of distributional treatment effects and documents the empirical relevance of crossing distributions in impact evaluations. Section 3 introduces the TEC together with the HDA and VDA frameworks for first-degree stochastic dominance and discusses their identification and causal interpretation. Section 4 derives the asymptotic inference for crossing points and local slopes, and Section 5 assesses its finite-sample performance in a Monte Carlo study. Section 6 develops the full synthetic-data workflow---TEC estimation, crossing-point inference, HDA, and VDA. Section 7 provides a short real-data diagnostic using PROGRESA. Section 8 contrasts the TEC with stochastic dominance, quantile regression, and causal-forest approaches, and Section 9 concludes.

\section{Recent advances in the analysis of Distributional Treatment Effects}\label{recent-advances-in-the-analysis-of-distributional-treatment-effects-dtes}
The study of DTEs is intimately related to the study of stochastic dominance. Early work on choice theory used stochastic dominance to address the need to compare distributions of outcomes rather than single moments such as means \citep{hadar1969,hanoch1969,rothschild1970}. Similarly, scholars working on social welfare theory showed how first- and second-order dominance criteria can be employed to make robust social welfare comparisons without specifying a particular utility function \citep{atkinson1970,shorrocks1983,levy1992,duclos2006}.

Subsequent econometric advances provided rigorous statistical procedures to test for dominance relations in finite samples. Methods developed by \citet{doksum1974}, \citet{bishop1989}, \citet{davidson2000}, \citet{barrett2003} and others introduced consistent, asymptotically valid tests for determining whether one distribution dominates another in the presence of sampling uncertainty. Collectively, this literature offers theoretically grounded and statistically robust methods for determining when one group, policy, or intervention leads to unambiguously better distributional outcomes.

It is thanks to these advances in stochastic dominance theory and the realization that ATEs could hide important distributional heterogeneities that the DTEs literature emerged. \citet{Abadie2002JASA} and \citet{AbadieAngristImbens2002Ecta} build on \citet{atkinson1970} and \citet{shorrocks1983} work to derive a quantile regression approach combined with an instrumental variable approach to study DTEs in a causal framework. Work by \citet{chernozhukov2004}, \citet{firpo2007}, \citet{rothe2010}, \citet{frolich2010}, \citet{maier2011}, \citet{firpo2015highlights}, \citet{callaway2018quantile}, \citet{chernozhukov2018generic}, and others further expanded this literature by working on inference and applications to different fields. These contributions showed that interventions often have heterogeneous effects that vary substantially across quantiles. Notable contributions in this vein include \citet{bitler2006}, who show that the mean impacts of welfare reform experiments mask substantial heterogeneity across the outcome distribution, and \citet{chernozhukov2013counterfactual}, who develop uniform inference procedures for counterfactual distribution functions that are closely related to inference on differences between CDFs.

More recently, a growing machine learning literature provides flexible, data-driven methods for estimating heterogeneous treatment effects in high-dimensional settings. Quantile regression forests recover the entire conditional distribution of outcomes rather than just its mean \citep{meinshausen2006}, causal forests estimate individual-level treatment effects with valid asymptotic inference \citep{wager2019}, and generalized random forests embed both in a common framework that solves a broad class of local moment conditions \citep{athey2019grf}. Subsequent refinements add robustness to high-dimensional nuisances and sharper local estimation \citep{oprescu2019,friedberg2020llf}. These methods supply adaptive, distribution-aware estimates of treatment effect heterogeneity that complement the discontinuity-focused tools we develop here.

While relatively recent, these contributions provide a robust and comprehensive set of tools to assess treatment effects all along the distribution of outcomes. In what follows, we aim to complement these methods by focusing on treatment effects discontinuities and provide specific tools to analyze these discontinuities.

\subsection{Crossing distributions in impact evaluations}\label{crossing-distributions-in-impact-evaluations}
The case for treatment-effect discontinuity analysis rests on an empirical premise: that the cumulative distribution functions of treated and control groups can genuinely cross in real impact evaluations, rather than one distribution simply dominating the other. The claim is not that crossings are ubiquitous. Some interventions generate approximate first-order dominance, while others generate heterogeneous but same-signed effects. However, in several well-known evaluations, a single policy lifts one part of the outcome distribution while leaving another unchanged or moving it in the opposite direction.

The clearest evidence comes from welfare reforms that combine work incentives with benefit limits. \citet{bitler2006} show that Connecticut's Jobs First program raised earnings at the bottom of the distribution, through its work incentive, while lowering them at the top, through its time limit and benefit withdrawal, producing a pronounced crossing of the earnings CDFs that the mean impact entirely conceals. The same authors document an analogous pattern for the Canadian Self-Sufficiency Project, where a time-limited earnings supplement left the lower half of the earnings distribution unchanged but raised its upper portion during the eligibility window, a crossing that appears and then dissolves once the supplement expires \citep{bitler2008}.

Education interventions provide a related example. In the randomized evaluation of a boarding school for disadvantaged students in France, \citet{behaghel2017} show that academic gains are delayed and concentrated among stronger students. The public replication data reproduce this pattern in the distribution of second-year mathematics scores. The assigned group is slightly below the control group at the bottom of the distribution, but above it from roughly the lower decile onward. This is not a dramatic crossing, but it illustrates that sign changes may be localized and substantively meaningful even when the average effect is positive.

Microcredit experiments provide a third source of sign-changing distributional evidence, although the patterns are noisier because profit outcomes are sparse and fat-tailed. In work on India, \citet{banerjee2015} report quantile treatment effect figures for informal borrowing and business profits. Their public replication files show negative or near-zero effects over much of the distribution and positive effects in the upper tail for some business-profit outcomes, especially among business owners at the second endline. The related Morocco and Mexico experiments \citep{crepon2015,angelucci2015} also find small average effects with larger gains concentrated among households more likely to operate or expand businesses. \citet{meager2019} and \citet{meager2022} aggregate seven microcredit experiments within a Bayesian hierarchical framework and show that near-zero average effects can obscure substantial heterogeneity across the distribution of outcomes. These studies therefore support the empirical relevance of sign-changing distributional effects, while also warning that crossings may be sensitive to outcome definition, sample restriction, and tail behavior.

The distinction between treatment heterogeneity and sign switch matters but understanding this distinction can be subtle. Two other accessible experimental datasets illustrate the point. The Kenya tracking experiment of \citet{duflo2011tracking} shows positive effects of tracking relative to non-tracking schools across the unconditional score distribution, while the regression-discontinuity comparison among marginal students is small and imprecise throughout. In the water-conservation experiment of \citet{ferraro2013nonpecuniary}, the social-norm treatments reduce water use for both low and high baseline users. The treatment effects differ in intensity, but they do not produce a substantively clear reversal across the outcome distribution. 

Taken together, evidence from welfare reform, education, and microcredit evaluations establishes that crossing distributions are not a theoretical curiosity but a recurring empirical possibility, and one for which conventional average-effect summaries are least informative. At the same time, the nearby counterexamples show why crossings must be diagnosed rather than assumed. It is precisely in these settings that the TEC, and the HDA and VDA tools developed below, are designed to be most useful. 

\section{Stochastic Dominance and the Treatment Effect Curve}\label{stochastic-dominance-and-the-treatment-effect-curve}
Let $X$ be a continuous random variable defined on $\mathbb{X}\subseteq\mathbb{R}$ that could be observed under two alternative treatments, say $A$ and $B$. Let
$X_{A,1},\ldots,X_{A,n_A}\overset{iid}{\sim} F_A$ and $X_{B,1},\ldots,X_{B,n_B}\overset{iid}{\sim} F_B$ be two independent samples, where $F_A$ and $F_B$ are absolutely continuous CDFs with corresponding probability density functions (pdfs) $f_A$ and $f_B$. From the classical theory of stochastic dominance, distribution $A$ is said to dominate distribution $B$ (at the first degree) if and only if
\begin{equation}\label{eq:sd_xspace}
F_{A}(x)\leq F_{B}(x) \quad \text{for all } x \in \mathbb{X},
\end{equation}
with at least one point $x$ at which the inequality is strict. Thus, in order to study DTEs and possible stochastic dominance relationships, we define the TEC as the difference between the two CDFs
\begin{equation}\label{eq:tec}
\Delta(x) = F_A(x) - F_B(x) \quad \text{for } x \in \mathbb{X}.
\end{equation}
Here $A$ and $B$ are generic labels. In the empirical applications below we set $A$ equal to the control or pre-treatment distribution and $B$ equal to the treated or post-treatment distribution, so that positive values of $\Delta(\cdot)$ indicate a rightward shift of the treated/post-treatment distribution.

The TEC provides a graphical and inferential device to assess first-order stochastic dominance and, when dominance fails, to locate the points where the two distributions cross and the marginal distributional effect changes sign. This curve has distinct advantages over classic dominance analysis and distributional impact evaluations based on stochastic dominance, quantile functions, quantile regressions, or causal random forests. TEC is intuitively simple, it is easy to plot, and it also provides a natural basis for inference on both marginal distributional treatment effects and crossing points ($\Delta(\cdot)$ is a continuous function but it can manifest discontinuity in sign). This curve is able to clearly separate areas where the treatment effect switches sign, even when dominance curves seem to overlap. This is arguably the main reason for using stochastic dominance in the context of impact evaluations and helps to distinguish areas with opposite effects, information that can be used for sub-group analysis.

When adopting TEC, at least two types of analysis become of interest. The first one, which we may call the ``Horizontal Discontinuity Analysis'' could focus on comparing sub-groups of observations located in outcome regions where the marginal distributional effect has opposite signs. Essentially, the sub-groups are divided by the points where the TEC crosses the zero line. This analysis could help understand why the distributional effect is positive in some regions and negative in others. For example, if a job training program had positive distributional effects among low-wage workers but negative effects among high-wage workers, HDA could help understand the factors explaining these different patterns.

Once the crossing points have been identified, the local slope of the TEC at each zero provides an estimate of the local marginal distributional
treatment effect, namely the rate at which the difference between the two CDFs changes as the outcome moves through the sign-switch point. Thus, the second analysis - which we may call the ``Vertical Discontinuity Analysis'' - could focus on the characteristics of observations just above or below the points where the TEC crosses the zero line. Similarly to the idea behind Regression Discontinuity Designs (RDD), one could think of these groups of observations as the key groups to observe for understanding why a marginal distributional effect switches sign. In fact, if CDFs cross multiple times resulting in several sign switches, TEC would be able to highlight multiple groups of observations situated just above and just below the zero-effect line. The focused analysis on these groups could provide useful information on where the treatment works or not, and for whom.

Thus, within this framework, the two main goals from an inferential point of view are to identify (i) the zeros - or crossing points - $x_0 \in \mathbb{X}$ such that $\Delta(x_0)=0$ (the horizontal problem underlying HDA) and (ii) the local slope $\Delta'(x_0)=f_A(x_0)-f_B(x_0)$ of the marginal distributional effect at those points (the vertical problem underlying VDA). We address these two problems in Section \ref{sec-inference}.
\begin{remark}\label{remRDD}
An important caveat distinguishes VDA from standard RDD. In conventional RDD the threshold is fixed exogenously by policy rules such as eligibility cutoffs or test-score requirements, whereas in VDA it is determined endogenously by the data, as the point where the estimated TEC crosses zero. VDA should therefore be viewed as discontinuity detection rather than formal RDD. We return to its identifying assumptions in Section \ref{sec-identification}. A second, distinct caveat is that a TEC crossing zero does not by itself imply a discontinuity in the underlying conditional average treatment effect (CATE) function. A CATE that declines smoothly through zero generates a TEC crossing with no jump at all. VDA should thus be understood as a test of whether the sign switch is abrupt, and the absence of a significant jump does not invalidate the crossing itself.
\end{remark}

In essence, and as compared to classic stochastic dominance analysis applied to impact evaluations, TEC is expected to be better at capturing treatment discontinuities and avoid the problem of blurring positive and negative treatment effects. It naturally lends itself to HDA and VDA analyses that can complement the distributional treatment work popularized with the quantile regression approach. Arguably, and from a policy perspective, distributional treatment analysis is the most meaningful as compared to ATEs when CDFs cross.

\subsection{Identification and Causal Interpretation}\label{sec-identification}
TEC represents the difference between the marginal cumulative distributions of potential outcomes. Under (conditional) ignorability, the two CDFs are identified from data and TEC can be estimated non-parametrically with standard tools (bootstrap, pointwise inference). In this sense, TEC requires no stronger identifying assumption than those used to identify marginal distributional treatment effects. This applies when comparing distributions of treated and non-treated individuals with cross-section data, or pre- and post-treatment distributions in panel data.

Formally, for every subject $i=1,\ldots,n$, let $(X_i(A),X_i(B))$ denote the potential outcomes under the two treatments, $Z_i$ is a vector of observable covariates and $W_i \in \{A,B\}$ denotes the treatment assignment. Assume that, for any $i$, $(X_i(A),X_i(B),W_i,Z_i)\overset{d}{=}(X(A),X(B),W,Z)$ and $0<\Pr(W_i=A\mid Z_i)<1$ a.s. Under the assumption of unconfoundedness (or selection on observables) $(X_i(A), X_i(B)) \perp W_i \mid Z_i$, so the marginal distributions are identified from the observed data as well as the TEC in (\ref{eq:tec}), where $F_A(x)=\Pr(X(A)\le x)$ and $F_B(x)=\Pr(X(B)\le x)$. While $\Delta(x)>0$ means that the treated marginal distribution is shifted rightward relative to the control marginal distribution at $x$. When TEC crosses zero, this indicates that the sign of the marginal distributional treatment effect changes at that point. 

However, the causal interpretation depends on the identifying assumptions. Under unconfoundedness with random treatment assignment, a crossing point at $x_0$ means that the treated and control marginal CDFs are equal at $x_0$ and (if $\Delta'(x_0)\neq 0$) they exchange their ordering locally. It does not imply that individuals with potential outcomes near $x_0$ have no treatment effects, nor does it identify the distribution of individual gains. Those interpretations would require stronger assumptions such as rank invariance or restrictions on the joint distribution of $(X_i(A),X_i(B))$.

When we use TEC crossings to define groups for HDA, comparing those above and below crossing points, we are comparing individuals located in regions where the marginal distributional effect has opposite signs. Under random assignment, differences between these groups in observable characteristics can shed light on effect heterogeneity. The causal forest approach then estimates CATEs within each group, which requires the same unconfoundedness assumption.

The VDA approach faces additional challenges because the thresholds are estimated rather than known ex ante. This introduces estimation uncertainty into the analysis and means that standard RDD asymptotic theory does not directly apply. We therefore recommend treating VDA results as suggestive of structural breaks rather than as formal causal estimates, and conducting sensitivity analyses with respect to bandwidth and threshold location.

\section{Asymptotic Inference for Crossing Points}\label{sec-inference}
The HDA and VDA frameworks introduced above hinge on the location of the points at which the TEC crosses zero and on the magnitude of its slope at those points. Both objects admit a fully developed asymptotic theory, with different optimal strategies for the two problems. 

\subsection{Locating crossing points}\label{locating-crossing-points}
To locate the zeros, the natural object is the \emph{un-smoothed} difference between the empirical CDFs (ECDFs), namely
\begin{equation}\label{eq:unsmoothdiff}
\hat{\Delta}(x) = \hat{F}_A(x) - \hat{F}_B(x)=\frac{1}{n_A}\sum_{i=1}^{n_A} \mathbf{1}\{X_{Ai}\le x\}-\frac{1}{n_B}\sum_{j=1}^{n_B} \mathbf{1}\{X_{Bj}\le x\},
\end{equation}
where $\mathbf{1}\{C\}$ denotes the indicator function of the event $C$. Since $\hat{\Delta}(x)$ is a step function, it can change sign only when $x$ crosses an observation from either sample. Therefore, to search for the zeros, it is sufficient to consider the ordered pooled sample $X_{(1)}\le X_{(2)}\le \cdots \le X_{(n)}$, with $n=n_A+n_B$. Because $\hat{\Delta}(\cdot)$ is constant between consecutive observations, each estimated zero is naturally located in an interval delimited by two consecutive observations of the pooled sample. The procedure thus reduces to finding a pair of consecutive outcomes such that $\hat{\Delta}(X_{(k)})\,\hat{\Delta}(X_{(k+1)})\le 0$ and by convention we set $\hat{x}_0 = (X_{(k)} + X_{(k+1)})/2$ (asymptotically, the error term derived from this approximation is negligible). In finite samples with $n_A=n_B$, $\hat{\Delta}(\cdot)$ changes in steps of $1/n_A$ and can equal zero exactly, occasionally over a short plateau of consecutive observations; in our implementation these zeros are treated as transparent, so that a crossing is identified as a genuine sign reversal between consecutive \emph{non-zero} values of $\hat{\Delta}(\cdot)$, with $\hat{x}_0$ again set to the midpoint of the two bracketing observations. This avoids recording a spurious crossing where $\hat{\Delta}(\cdot)$ merely touches zero without changing sign, and it coincides with the convention above whenever $\hat{\Delta}(\cdot)$ does not vanish on the lattice. Whereas, if $\hat{\Delta}(X_{(k)})\,\hat{\Delta}(X_{(k+1)}) > 0$ for every consecutive pair, no crossing is detected and the empirical TEC is one-signed over the observed support, which is consistent with first-order dominance of one distribution over the other.

Assuming that the set of crossing points is non-empty, we impose the following regularity conditions on each zero:

\noindent \textbf{C1.} $x_0$ is a non-degenerate and isolated crossing, namely it lies in the interior of $\mathbb{X}$ and, within a neighborhood of $x_0$, is the only point at which the two CDFs cross.\\

\noindent \textbf{C2.} $x_0$ is non-tangential, namely $\Delta'(x_0) = f_A(x_0) - f_B(x_0) \neq 0$.\\
Condition C1 is standard and widely satisfied in economic applications, where zeros at the boundaries of the support are usually of limited interest and the presence of two extremely close crossing points has little substantive interpretation. Condition C2 guarantees local identification. It implies that, in a neighborhood of $x_0$, $\Delta(x) = \Delta'(x_0)(x - x_0) + o(|x-x_0|)$, so that small vertical perturbations of $\Delta(\cdot)$ translate into controlled horizontal perturbations of the zero. By contrast, if $\Delta'(x_0) = 0$ the crossing is locally flat, small stochastic fluctuations of the empirical process may induce large errors in the estimated crossing point, and no significant treatment effect exists at that point.

These two conditions define two complementary diagnostic steps that we use throughout. The first step concerns the existence and location of an isolated interior crossing. Operationally, we detect candidate crossings by searching for sign reversals of $\hat\Delta(\cdot)$ on the ordered pooled sample, and then attach the Wald confidence interval derived below to each detected crossing. In the Monte Carlo study, the frequency of spurious detections under a no-crossing design provides the false-positive rate of this detection rule, while the frequency of correct detections when crossings are present provides its detection power. Then, we ask whether the located crossing is non-tangential, testing $H_0:\Delta'(x_0)=0$ against $H_1:\Delta'(x_0)\neq 0$. A crossing must first be detected and located before its steepness can be assessed: this test 
is used to assess the non-tangentiality of the previously identified crossing point which, in turn, amounts to testing whether the local marginal distributional treatment effect is significantly different from zero.

Under C1-C2 and the absolute continuity of the CDFs, the pdfs $f_A$ and $f_B$ are twice continuously differentiable and admit a Taylor expansion in a neighborhood of $x_0$. Letting $p_0 = F_A(x_0) = F_B(x_0)$ and $a_n = n_A^{-1} + n_B^{-1}$, \citet{hawkins1991} show that, as $n_A, n_B \to \infty$,
\begin{equation}\label{eq:asnorm}
a_n^{-1/2}(\hat{x}_0 - x_0) \;\overset{d}{\hookrightarrow}\; N\!\left(0,\; \dfrac{p_0(1-p_0)}{[\Delta'(x_0)]^2}\right),
\end{equation}
so that
\begin{equation}\label{eq:asnorm1}
\mathrm{Var}(\hat{x}_0) \approx a_n \left\{\dfrac{p_0(1-p_0)}{[\Delta'(x_0)]^2}\right\}.
\end{equation}
The accuracy of the crossing-point estimator therefore increases with the magnitude of the local slope $|\Delta'(x_0)|$. The steeper the crossing, the more sharply the zero is identified. Crossings associated with stronger treatment effects are easier to locate with high precision. Conversely, if $f_A(x_0)\simeq f_B(x_0)$, the crossing is nearly flat and the uncertainty in its location increases sharply.

Within this setting, plug-in estimates are typically employed,
\begin{equation}\label{eq:plugin}
\hat{p}_0 = \frac{\hat{F}_A(\hat{x}_0) + \hat{F}_B(\hat{x}_0)}{2} \qquad \text{and} \qquad \widehat{\mathrm{Var}}(\hat{x}_0) = a_n \frac{\hat{p}_0(1-\hat{p}_0)}{[\hat{\Delta}'(\hat{x}_0)]^2},
\end{equation}
so that the Wald-type $(1-\alpha)$ confidence interval for $x_0$ is $\left[\hat{x}_0 \pm z_{1-\alpha/2}\sqrt{\widehat{\mathrm{Var}}(\hat{x}_0)}\right]$.
 %and the Wald statistic for testing $H_0: x_0 = \mathrm{x}$ is simply given by $W = \widehat{\mathrm{Var}}(\hat{x}_0)^{-1/2}(\hat{x}_0 -\mathrm{x})$.

\subsection{Estimating the local slope}\label{estimating-the-local-slope}
Having estimated the number of crossing points of the TEC and located $\hat{x}_0$, the local slope $\Delta'(x_0) = f_A(x_0) - f_B(x_0)$ could be naturally estimated by kernel smoothing of the two densities at that point (see, e.g., \cite{parzen1962}). Let $K(\cdot)$ be a positive and symmetric kernel (e.g., Gaussian) and $h > 0$ a bandwidth, then the kernel estimates of the two pdfs are
\begin{equation}\label{eq:kde}
\hat{f}_{Ah}(x) = \frac{1}{h\,n_A}\sum_{i=1}^{n_A} K\!\left(\frac{x - X_{Ai}}{h}\right) \qquad \text{and} \qquad \hat{f}_{Bh}(x) = \frac{1}{h\,n_B}\sum_{j=1}^{n_B} K\!\left(\frac{x - X_{Bj}}{h}\right)
\end{equation}
and the natural slope estimator is $\hat{\Delta}'_h(x_0) = \hat{f}_{Ah}(\hat{x}_0) - \hat{f}_{Bh}(\hat{x}_0)$. Let $\mu_2(K) = \int u^2 K(u)du$ and $R(K) = \int K(u)^2du$, then 
\begin{equation}\label{eq:kdebias}
E\!\left\{\hat{f}_{jh}(x_0)\right\} = f_j(x_0) + \frac{\mu_2(K)\,f_j''(x_0)\,h^2}{2} + o(h^2), \qquad j = A, B
\end{equation}
and therefore
\begin{equation}\label{eq:biasdelta}
E\!\left\{\hat{\Delta}'_h(x_0) - \Delta'(x_0)\right\} = \frac{\mu_2(K)}{2}\bigl[f_A''(x_0) - f_B''(x_0)\bigr]\, h^2 + o(h^2);
\end{equation}
from the independence of the two samples,
\begin{equation}\label{eq:slopevar}
\mathrm{Var}\bigl(\hat{\Delta}'_h(x_0)\bigr) = \frac{R(K)}{h}\!\left(\frac{f_A(x_0)}{n_A} + \frac{f_B(x_0)}{n_B}\right) + o\bigl((h a_n)^{-1}\bigr),
\end{equation}
which can be consistently estimated by
\begin{equation}\label{eq:slopevarhat}
\widehat{\mathrm{Var}}\bigl(\hat{\Delta}'(x_0)\bigr) = \frac{R(K)}{h}\!\left(\frac{\hat{f}_{Ah}(\hat{x}_0)}{n_A} + \frac{\hat{f}_{Bh}(\hat{x}_0)}{n_B}\right).
\end{equation}
Locally at $x_0$, the asymptotic MSE of $\hat{\Delta}'_h$ is therefore
\begin{equation}\label{AMSE}
\mathrm{AMSE}\bigl(\hat{\Delta}'_h(x_0)\bigr) \approx \frac{\mu_2(K)^2}{4}\bigl[f_A''(x_0) - f_B''(x_0)\bigr]^2 h^4 + \frac{R(K)}{h}\!\left(\frac{f_A(x_0)}{n_A} + \frac{f_B(x_0)}{n_B}\right);
\end{equation}
thus, if $f_A''(x_0) \neq f_B''(x_0)$, the optimal bandwidth $h^*$ is of order $O(n^{-1/5})$ (provided that $n_A \simeq n_B$) and depends on $x_0$. A standard plug-in implementation based on a pilot estimate of $x_0$, $f_j(x_0)$ and $f_j''(x_0)$ follows \citet{sheather1991}. Whereas, when $f_A''(x_0) = f_B''(x_0)$ the leading bias term vanishes and higher-order Taylor expansions are needed.
\begin{remark} 
\emph{Why use the un-smoothed ECDFs to locate crossings}. Let $\hat{\Delta}_h$ denote the smoothed version of the TEC obtained by integrating $\hat{f}_{jh}$ for $j=A,B$ and let $\hat{x}_{0h}$ be the corresponding estimated crossing point. By a Taylor expansion around $x_0$,
$0 = \hat{\Delta}_h(\hat{x}_{0h}) \approx \hat{\Delta}_h(x_0) + \Delta'(x_0)(\hat{x}_{0h} - x_0)$, so that
\begin{equation}\label{eq:smoothbias}
\hat{x}_{0h} - x_0 \approx (\Delta(x_0) - \hat{\Delta}_h(x_0))/\Delta'(x_0).
\end{equation}
Hence any vertical bias introduced by smoothing the TEC at $x_0$ induces a horizontal bias in the estimated crossing point. From (\ref{eq:asnorm1}), $\sqrt{\mathrm{Var}(\hat{x}_0)} = O(a_n^{1/2})$, while the standard kernel bias in (\ref{eq:biasdelta}) is of order $O(h^2)$. Making the horizontal bias asymptotically negligible therefore requires $h = o(a_n^{1/4})$, which is much smaller than the standard density-estimation bandwidth $h = O(n^{-1/5})$ (see \cite{wand1995}). Thus, the amount of smoothing that can be tolerated for locating the zeros is asymptotically very limited, which justifies the use of the un-smoothed ECDFs to locate the crossings.
\end{remark}
\begin{remark}\emph{Evaluating the slope at $\hat{x}_0$ rather than at $x_0$.} Since the slope is evaluated at the estimated crossing point $\hat{x}_0$, the relevant quantity is $\mathrm{AMSE}\bigl(\hat{\Delta}'_h(\hat{x}_0)\bigr)$ rather than $\mathrm{AMSE}\bigl(\hat{\Delta}'_h(x_0)\bigr)$ in (\ref{AMSE}). However, under suitable regularity conditions, these two quantities are asymptotically equivalent. Indeed, from (\ref{eq:asnorm}), $\hat{x}_0 - x_0 = O_p(a_n^{1/2})$, while the kernel density estimator is local over a window of width $h$. The localization error is thus negligible at the scale of the kernel estimator as long as $a_n^{1/2} = o(h)$. For instance, if $n_A \simeq n_B$, then $a_n \simeq n^{-1}$ and the choice $h \simeq n^{-1/5}$ satisfies this requirement, since $n^{-1/2} = o(n^{-1/5})$. Evaluating the slope estimator at $\hat{x}_0$ (rather than at $x_0$) is therefore asymptotically innocuous.
\end{remark}
For instance, take the Gaussian kernel $K \equiv \phi(\cdot)$ to be the pdf of the standard normal, so that $\mu_2(K) = 1$ and $R(K) = (4\pi)^{-1/2}$. Thus,
\begin{equation}\label{eq:amsegauss}
\mathrm{AMSE}\bigl(\hat{\Delta}'_h(x_0)\bigr) \approx \frac{1}{4}\bigl(f_A''(x_0) - f_B''(x_0)\bigr)^2 h^4 + \frac{1}{2h\sqrt{\pi}}\!\left(\frac{f_A(x_0)}{n_A} + \frac{f_B(x_0)}{n_B}\right)
\end{equation}
and the optimal bandwidth $h^*(x_0)$ is
\begin{equation}\label{eq:hopt}
h^*(x_0) = \left[\frac{\dfrac{f_A(x_0)}{n_A} + \dfrac{f_B(x_0)}{n_B}}{2\sqrt{\pi}\,\bigl[f_A''(x_0) - f_B''(x_0)\bigr]^2}\right]^{1/5},
\end{equation}
which is of order $O(n^{-1/5})$ (provided that $f_A''(x_0) \neq f_B''(x_0)$) and depends on $x_0$. Pilot studies can be employed to obtain plug-in estimates of $h^*(x_0)$.

When multiple crossing points exist, the procedure above can be applied locally to each of them, provided that they are isolated. Otherwise the local kernel estimates would be based on largely overlapping neighborhoods and the estimated slopes would be strongly dependent. A common bandwidth $h$ for both $f_A$ and $f_B$ is a natural and stable default. In the presence of a-priori information about different curvatures of the two densities, allowing for different bandwidths may improve flexibility.

At any fixed evaluation point the slope estimator is asymptotically normal, so the non-tangentiality condition C2 can be verified directly. To test $H_0:\Delta'(x_0)=0$, a locally flat (tangential) crossing, against $H_1:\Delta'(x_0)\neq 0$, the Wald statistic could be employed. At each $\hat x_0$, from (\ref{eq:biasdelta}) the bias of $\widehat{\Delta}'_h(\hat x_0)$ is of order $h^2$: by estimating the leading kernel bias via
\begin{equation}\label{eq:biashat}
\widehat B_h(\hat x_0)=\frac{\mu_2(K)}{2}\left[\hat f_A''(\hat x_0)-\hat f_B''(\hat x_0)\right]h^2,
\end{equation}
the bias-corrected Wald statistic is
\begin{equation}\label{eq:waldC2}
W^{bc}=\frac{\widehat{\Delta}'_h(\hat x_0)-\widehat B_h(\hat x_0)}{\sqrt{\widehat{\operatorname{Var}}\left\{\widehat{\Delta}'_h(\hat x_0)\right\}}}.
\end{equation}
Under $H_0$, $W^{bc} \overset{d}{\hookrightarrow} N(0,1)$, so the null of a locally flat crossing is rejected at level $\alpha$ whenever
$\left|W^{bc}\right|>z_{1-\alpha/2}$.
The location step asks \emph{where} a crossing lies, while the test asks \emph{whether} the marginal distributional effect has a non-zero local slope there. Rejecting $H_0$ provides evidence that the detected crossing is non-tangential and locally well identified (while failing to reject $H_0$ points to a flat crossing, whose location is poorly identified).

Alternatively, the classical Wald statistic could also be applied without bias-correction, but this choice requires undersmoothing since the kernel bias of $\hat{\Delta}'_h$ should vanish faster than the standard deviation, namely $h^2 = o\bigl((a_n/h)^{1/2}\bigr)$; if $n_A \simeq n_B$, then $a_n$ is of order $n^{-1}$ and $h^2 = o\bigl((nh)^{-1/2}\bigr)$, i.e. $h^2\sqrt{nh}\rightarrow 0$; thus, the Silverman rule-of-thumb bandwidth ($h = O(n^{-1/5})$) is not enough: a more undersmoothed choice, such as half of Silverman's rule of thumb, restores calibration; locally tuned plug-in bandwidths in the spirit of \citet{sheather1991} are an alternative. And because $\hat{x}_0$ is estimated from the same sample, the test is asymptotically valid at a pre-specified threshold and is best read as a diagnostic when evaluated at an estimated crossing.

\section{A Monte Carlo Study}\label{sec-montecarlo}
We assess the finite-sample performance of the inference procedures of Section \ref{sec-inference} with a Monte Carlo study. The study addresses four questions. How reliably does the crossing-detection rule detect crossing points? How often does it flag spurious crossings when none exist (its false-positive rate, measured under the placebo design)? How accurate are the point estimator and the Wald confidence intervals? And how much does the explicit bias correction of the local slope improve slope recovery and the calibration of the crossing-location interval?

We consider four data generating processes (DGPs) on $\mathbb{X}=\mathbb{R}^{+}$. While sample $A$ is always drawn from a standard log-normal distribution (i.e., $\ln X_A \sim N(0,1)$), the scenarios differ in the distribution of sample $B$: (i) a \emph{placebo} case mimicking a homogeneous treatment effect, with $\ln X_B \sim N(0.25, 1)$, where a pure location shift in logs induces a first-order stochastic dominance of $B$ over $A$ (the TEC has no interior crossing points); (ii) a \emph{steep single-crossing} scenario with $\ln X_B \sim N(0.2, 1.5^2)$, which yields one crossing at $x_0 = 0.670$, with $p_0 = 0.345$ and slope $\Delta'(x_0) = 0.183$; (iii) a \emph{flat single-crossing} case with $\ln X_B \sim N(0.05, 1.1^2)$, with one crossing at $x_0 = 0.607$ ($p_0 = 0.309$) and a slope roughly three and a half times smaller, $\Delta'(x_0) = 0.053$; and (iv) a \emph{double-crossing} scenario in which $\ln X_B$ is a two-component normal mixture, $0.5\,N(-0.8, 0.4^2) + 0.5\,N(1.2, 0.4^2)$, producing crossings at $x_0 = 0.376$ (with $p_0 = 0.164$ and slope $-0.543$) and $x_0 = 0.967$ (with $p_0 = 0.487$ and slope $0.326$), a pattern similar to the synthetic-data illustration in the next section. True crossing points and slopes are computed from the exact CDFs by root finding.

For each design we draw 1,000 replications of two independent samples of equal size $n_A=n_B \in \{500;\, 2{,}000;\, 8{,}000\}$ (so that the pooled sample has size $n \in \{1{,}000;\, 4{,}000;\, 16{,}000\}$) and apply exactly the procedures described in Section \ref{sec-inference}. Crossings are located through sign reversals of the un-smoothed ECDF difference on the pooled sample, restricted to the interior ($0.05 < \hat{p}_0 < 0.95$) since crossings at the boundaries of the support $\mathbb{X}$ are of limited substantive interest and with nearby sign flickers merged. Local slopes are estimated with Gaussian kernels at Silverman's rule-of-thumb bandwidth and bias-corrected with the explicit plug-in adjustment of Section \ref{sec-inference}, while 95\% Wald confidence intervals for the location have been used. Bias, RMSE, standard error accuracy, and coverage are computed conditional on the replication detecting the true number of interior crossings. Table \ref{tbl-mc} reports the results.

[Table 1 approximately here.]

Four findings stand out. First, detection power depends critically on the steepness of the crossing, exactly as equation (\ref{eq:asnorm1}) predicts. The steep crossing is found in 94\% of replications already at $n=1{,}000$ and virtually always for $n \geq 4{,}000$, while the flat crossing - whose slope is 3.5 times smaller - is found in only 22\% of replications at $n=1{,}000$ and 86\% at $n=16{,}000$. Researchers should expect crossings associated with weak local slopes of the distributional effect to require large samples to be detected reliably. Second, the placebo design shows that spurious crossings are a small-sample phenomenon. The false positive rate is 14\% at $n=1{,}000$ but essentially zero for $n \geq 4{,}000$. Third, conditional on detection, the point estimator is approximately unbiased (the bias is an order of magnitude smaller than the RMSE) and converges at the parametric rate. Quadrupling the sample size halves the RMSE, consistent with the $\sqrt{n}$-consistency in equation (\ref{eq:asnorm}). The relative precision across designs also matches the theory. At $n=16{,}000$ the RMSE of the flat crossing is about 3.2 times that of the steep crossing, close to the inverse ratio of the slopes. Fourth, the Wald intervals are conservative for the steep and double crossings, where empirical coverage ranges from 93\% to 100\%, but the flat crossing under-covers in small samples (87\% at $n=1{,}000$, rising to 94\% at $n=16{,}000$ as the asymptotic approximation improves).

These patterns trace back to the estimated standard error in equation (\ref{eq:asnorm1}), which varies inversely with the local slope. Even after bias correction, the slope is hardest to pin down where the density is sparse or changing rapidly. At the first crossing of the double scenario, located in the lower tail ($p_0 = 0.16$), the slope is understated and the standard error is correspondingly inflated, exceeding the Monte Carlo standard deviation by a factor of about two to nine (the SE ratio falls from $8.7$ at $n=1{,}000$ to $2.1$ at $n=16{,}000$), which is why the interval over-covers. At the flat crossing, by contrast, the small slope is estimated with relatively more noise and the asymptotic approximation bites only slowly, so the standard error is mildly understated (SE ratio $0.83$ to $0.94$) and the interval under-covers until $n$ is large. This supports the recommendation of Section \ref{sec-inference} to tune the bandwidth to the local slope, for instance with the plug-in choice of \citet{sheather1991}, when the slope itself is the object of interest. Importantly, the crossing-point estimator is unaffected, since it is located from the un-smoothed ECDF difference and uses no smoothing.

[Table 2 approximately here.]

Table \ref{tbl-mcbw} isolates the role of the bias correction at the Silverman bandwidth (steep design, $n=4{,}000$). The conventional Gaussian slope estimator recovers only 67\% of the true slope, which inflates the estimated standard error (SE ratio $1.50$) and makes the Wald interval conservative (99.7\% coverage). The explicit plug-in bias correction of Section \ref{sec-inference} raises the recovered slope to 93\% of its true value, brings the SE ratio close to one ($1.05$), and delivers near-nominal coverage (96.6\%). Here the SE ratio and coverage refer to the crossing-location interval, whose width depends on the slope plug-in; removing the kernel bias therefore improves both slope recovery and location-interval calibration, whereas leaving it uncorrected trades calibration for conservatism. This exercise speaks to slope recovery and the calibration of the location interval; the size and power of the C2 slope test itself are not simulated here. It is also distinct from the crossing-location problem in Remark 2 of Section \ref{sec-inference}: the location of $\hat{x}_0$ comes from the un-smoothed ECDF difference, while the bias correction affects only the slope estimate that enters the standard error.

Three limitations of the study should be noted. The designs are based on log-normal families with equal sample sizes, and performance may differ for distributions with heavier tails or strongly unbalanced samples. Coverage statistics are computed conditional on detecting the correct number of crossings, which induces a mild selection effect, most relevant for the flat scenario with small $n$. And the study evaluates the inference procedures of Section \ref{sec-inference} only. The properties of the downstream HDA and VDA analyses, which inherit estimated rather than known thresholds, are left for future work.

\section{A Simulation with Synthetic Data}\label{a-simulation-with-synthetic-data}
The Monte Carlo study of Section \ref{sec-montecarlo} validated the statistical properties of the crossing-point inference under known data generating processes. This section is the full worked example of the toolkit - i.e., TEC estimation, crossing-point inference, HDA and VDA - in a controlled setting where crossings are steep, well separated and known by construction. Section \ref{sec-progresa} uses a real randomized experiment as a diagnostic check, showing both distributional dominance and weak spillover sign changes that are too flat to justify threshold-based HDA or VDA.

We use a synthetic dataset provided by the World Bank for the purpose of simulating survey data.\footnote{The dataset can be freely downloaded from: https://microdata.worldbank.org/index.php/catalog/5906.} This dataset has been generated from hundreds of real microdata sets and includes 2.5 million household observations accounting for approximately 10 million individuals. For simulation purposes, this is considered a census. From this synthetic census, we drew a simple random sample of households. After trimming a small number of extreme per capita expenditure values, the analysis sample comprises 2,420 household observations and 9,849 individual observations. This is the dataset used for our simulations and inference.

Next, we randomly assign treatment to 50\% of respondents by attributing an extra two years of education, together with mild behavioral responses in savings and investment. The baseline (pre-treatment) values of education, savings, and investment serve as covariates in the analyses below. The outcome variable (per capita expenditure) was manipulated to produce a nuanced treatment effect pattern. Individuals below the 30th percentile of baseline expenditure experience mild short-term declines, middle-income individuals (30th-60th percentile) experience significant gains, and upper-income individuals experience muted or slightly negative outcomes. These changes produce a post-treatment CDF that crosses the pre-treatment CDF twice. The effect-scale boundaries are set at the 30th and 60th percentiles of baseline expenditure but, because the imposed effects are multiplicative shifts, the second crossing of the CDFs occurs at roughly the 70th percentile of the distribution. Because the post-treatment data are generated from the same pre-treatment sample, the data form a panel and our main exercise compares pre- and post-treatment distributions (the ``pre-post'' comparison). This pre-post estimand should be read as the distributional change in the full synthetic population under the imposed treatment process, not as a pure treated-versus-control contrast. Since treatment was assigned randomly, a comparison of treated and non-treated individuals post-treatment (the ``control-treated'' comparison) is also available. We use it as an independent-samples check in the crossing-point inference below, where the asymptotic theory of Section \ref{sec-inference} is directly applicable.

To quantify sampling uncertainty of both the empirical CDFs and the TEC, we construct pointwise nonparametric confidence intervals using bootstrap re-sampling. For each distribution being estimated, we generated 999 bootstrap samples drawn with replacement from the original sample. For every bootstrap replicate, we recomputed the empirical CDF over a fixed grid of expenditure values. The resulting bootstrap distribution of CDF values at each grid point was then used to form 95\% confidence intervals defined as the 2.5th and 97.5th percentiles across bootstrap draws. Because pointwise bands invite multiple-testing concerns when the curve is scanned for significant regions, we complement them with uniform (sup-norm) 95\% confidence bands for the TEC, computed from the bootstrap distribution of the maximal absolute deviation of the TEC across the grid. Bootstrap re-sampling is performed at the individual level, re-sampling pairs of pre- and post-treatment observations jointly to preserve their within-person dependence structure. This approach provides a fully nonparametric, distribution-free estimate of uncertainty around distributional treatment effects.

\begin{figure}[htbp]\centering
\includegraphics[width=\textwidth]{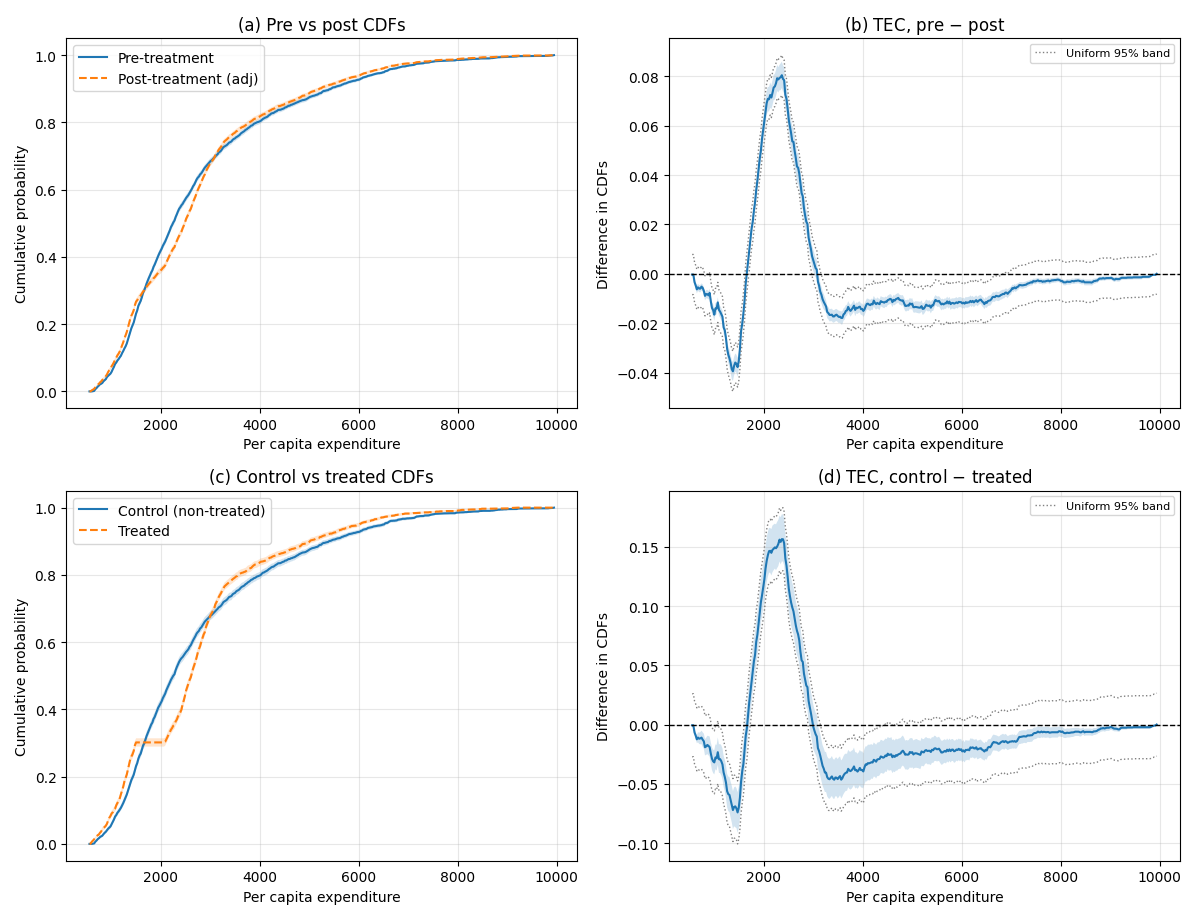}
\caption{Cumulative distribution functions (CDFs) and Treatment Effects Curves (TECs) for the synthetic data, under two comparisons. Top row reports the pre-post comparison. Panel (a) shows the pre- and post-treatment CDFs and panel (b) the corresponding TEC $\Delta(x)=F_{\text{pre}}(x)-F_{\text{post}}(x)$. Bottom row reports the control-treated comparison among post-treatment units. Panel (c) shows the control (non-treated) and treated CDFs and panel (d) the corresponding TEC $\Delta(x)=F_{\text{control}}(x)-F_{\text{treated}}(x)$. Shaded regions are pointwise 95\% bootstrap confidence intervals (999 replications), dotted lines are uniform (sup-norm) 95\% bands, and the dashed horizontal line marks zero, where the sign of the marginal distributional effect changes. The pre-post TEC band uses paired resampling, while all marginal CDF bands are resampled separately; the control-treated TEC band uses independent resampling.}\label{fig-CDFTECs}
\end{figure}

Figure~\ref{fig-CDFTECs} shows the CDFs and the corresponding TEC for the pre-post comparison in panels (a) and (b) and for the control-treated comparison in panels (c) and (d). While the two CDFs in panel (a) appear almost indistinguishable to the naked eye, the TEC in panel (b) cleanly separates three groups of observations: a group with negative effects at the lower end of the distribution, a group with positive effects around the lower-middle, and a group with mild negative effects in the upper part. This is by construction, but it illustrates the first practical advantage of the TEC. Treatment discontinuities that are nearly invisible in standard CDF comparisons become immediately apparent. Panels (c) and (d) show that the control-treated comparison, which contrasts independent groups of treated and non-treated units after treatment, reproduces the same three-group pattern with a larger amplitude, confirming that the structure is not an artifact of the within-person dependence of the pre-post panel. 

[Figure 1 approximately here.]

\subsection{Inference on the Crossing Points}\label{inference-on-the-crossing-points}
Before turning to the discontinuity analyses, we apply the crossing-detection rule of Section \ref{sec-inference} to detect and locate the crossing points of the TEC and to quantify their sampling uncertainty. Crossings are located by scanning the ordered pooled sample for sign changes of the un-smoothed ECDF difference. The local slope at each crossing is estimated with Gaussian kernel density estimates at a common Silverman rule-of-thumb bandwidth (of order $n^{-1/5}$), with the kernel bias removed by the explicit plug-in correction of Section \ref{sec-inference}, so that the bias-corrected Wald test of equation (\ref{eq:waldC2}) is asymptotically valid. Standard errors for the location follow from the asymptotic variance in equation (\ref{eq:asnorm1}), and the non-tangentiality Wald test follows equation (\ref{eq:waldC2}). We restrict attention to interior crossings $(0.05 < \hat{p}_0 < 0.95$), since crossings at the boundaries of the support are of limited substantive interest. Table \ref{tbl-crossings} reports the results.

[Table 3 approximately here.]

For the pre-post comparison, the TEC crosses zero at per capita expenditure of approximately 1,658 $(\hat{p}_0 = 0.30$; 95\% CI [1,589; 1,726]) and 3,079 $(\hat{p}_0 = 0.70$; 95\% CI [2,931; 3,227]). The control-treated comparison yields very similar estimates (1,669 and 2,996). The estimated slopes are positive at the first crossing and negative at the second, consistent with treatment effects switching from negative to positive and back to negative along the distribution, and the C2 test rejects a flat crossing at every point ($p<0.001$), confirming non-tangential crossings. The first crossing is estimated more precisely than the second because the TEC is steeper there, in line with equation (\ref{eq:asnorm1}). Two caveats apply. First, the asymptotic theory of Section \ref{sec-inference} assumes independent samples. This holds for the control-treated comparison, while for the pre-post comparison the within-person dependence of the panel means the reported intervals and $p$-values should be read as approximations. Second, the C2 test uses the explicit plug-in bias correction of Section \ref{sec-inference} at the Silverman bandwidth, which removes the leading kernel bias. A fully data-driven plug-in bandwidth in the spirit of \citet{sheather1991} is a refinement left for the practitioner. The estimated pre-post crossing points are saved and used as thresholds in all the HDA and VDA analyses that follow.

\subsection{A Horizontal Discontinuity Analysis}\label{a-horizontal-discontinuity-analysis-hda}
An HDA can add insights to this picture. It compares groups of observations delimited by the points where the curve crosses the zero line. In our example, these points delimit three groups which we can compare to search for insights on factors that could determine a switch in sign. As an illustration, we focus on the pre-post treatment exercise with the crossing points established in the previous section at 1,658 $(\hat{p}_0 = 0.30$; 95\% CI [1,589; 1,726]) and 3,079 $(\hat{p}_0 = 0.70$; 95\% CI [2,931; 3,227]).

\begin{figure}[htbp]\centering
\includegraphics[width=\textwidth]{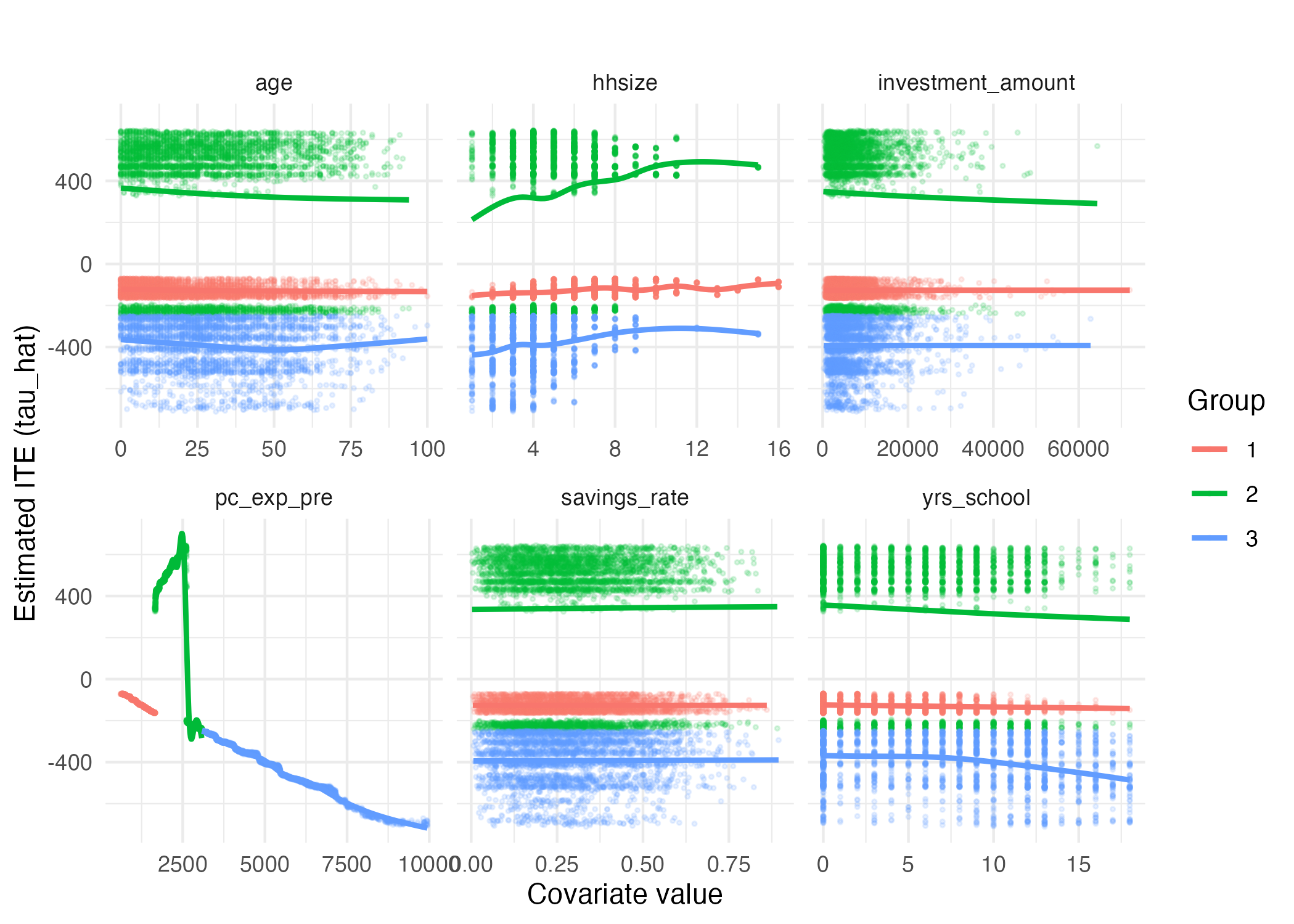}
\caption{Treatment effects across covariates by group, causal forest estimates (synthetic data).}\label{fig-TECov}
\end{figure}

We explore the determinants of treatment effects across the three groups divided by the two crossing points. This can be done with a variety of models. Here we use causal random forest \citep{athey2019grf,wager2019} fitting one model for each group identified by the TEC. We estimate how the conditional treatment effect varies within and across these groups. For each group, we model the causal effect of the treatment on post-treatment expenditure while controlling for socio-economic and behavioral covariates. We omit variable importance and individual treatment effects by group and show instead estimated treatment effects across covariates by group (Figure~\ref{fig-TECov}). Pre-treatment per capita expenditure is the main driver of results together with household size, investments, and years of education, all correlates of per capita expenditure. We can also observe a certain heterogeneity of effects within groups, particularly across group 2, which is the only group located in the positive region of the TEC. One caveat is in order. Because the grouping variable is defined from pre-treatment expenditure, which is also the most important covariate in the forests, group membership is endogenous to the running variable. The sample-splitting built into the causal forest estimator mitigates overfitting concerns, but within-group results should be read as descriptive of effect heterogeneity rather than as causal estimates in their own right.

Because the data are synthetic, these group differences are imposed by construction rather than estimated. With real data, the same output would invite a behavioral interpretation of why low-, middle-, and high-expenditure households respond through different channels, which is exactly the question the framework is designed to raise.

[Figure 2 approximately here.]

\subsection{A Vertical Discontinuity Analysis}\label{a-vertical-discontinuity-analysis-vda}
A Vertical Discontinuity Analysis applied to a TEC is an RDD-inspired local discontinuity analysis centered on the points where the TEC curve crosses the zero line. In our previous example, these are the points where per capita expenditure is equal to 1,658 and 3,079 (or the corresponding quantiles of 0.30 and 0.70, as estimated above). Using the observed individual-level change $\Delta_i = \text{pc\_exp\_post\_adj}_i - \text{pc\_exp\_pre}_i$ as the outcome and the pre-treatment expenditure as the running variable, local linear discontinuity analyses are estimated with triangular kernel weighting in narrow bandwidths around each threshold. This approach tests whether the observed change exhibits a statistically significant upward or downward jump at the points that separate the three distributional-effect regions, thereby linking the distributional heterogeneity to local discontinuities in responses. It is worth stressing that the individual-level change $\Delta_i$ is observable here only because the synthetic data form a panel in which the same individuals are observed pre- and post-treatment. In cross-section designs comparing treated and control units, individual effects are not observed and the VDA outcome must be replaced by estimated individual treatment effects (for example, from a causal forest), which introduces an additional layer of estimation error into the analysis.

As discussed in Remark \ref{remRDD}, the VDA differs fundamentally from standard RDD in that the thresholds are estimated from the data rather than fixed by exogenous policy rules. Standard RDD asymptotic theory, which assumes a known threshold, does not directly apply and results should be read as evidence of structural breaks rather than sharp causal discontinuities. The simulation is a useful illustration of the abrupt-versus-gradual distinction noted earlier. The local discontinuity analysis detects a significant jump at the first boundary, where the data-generating process imposes a discrete effect-scale step, but no significant jump at the second crossing, which falls away from the imposed step. In real data sign switches may likewise be smooth, in which case the analysis would correctly find no jump even though the crossing is real. 

\begin{figure}[htbp]\centering
\begin{subfigure}[t]{0.48\textwidth}\centering\includegraphics[width=\textwidth]{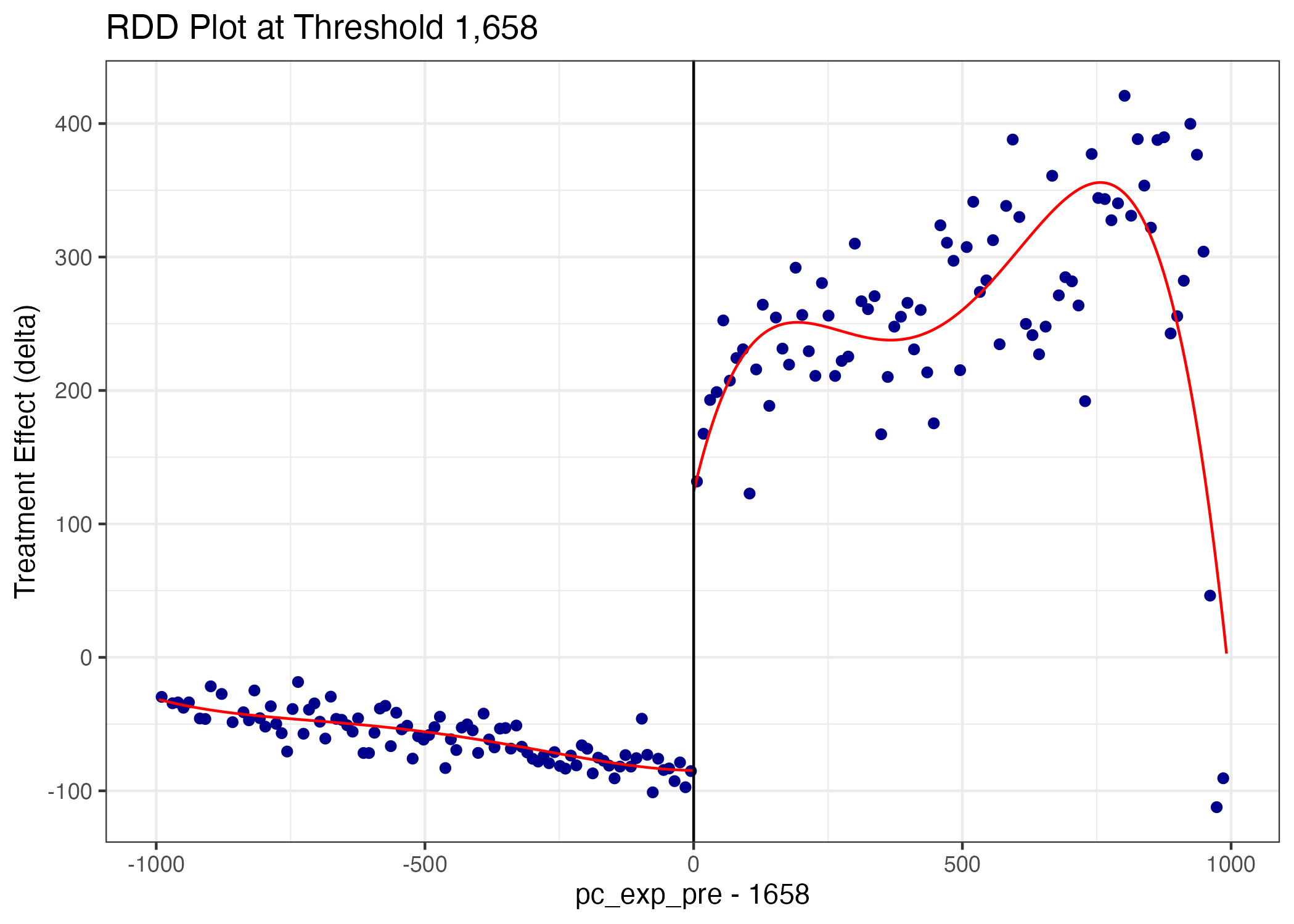}
\caption{RDD plot at the first estimated crossing point}\label{fig-RDD1}\end{subfigure}\hfill
\begin{subfigure}[t]{0.48\textwidth}\centering\includegraphics[width=\textwidth]{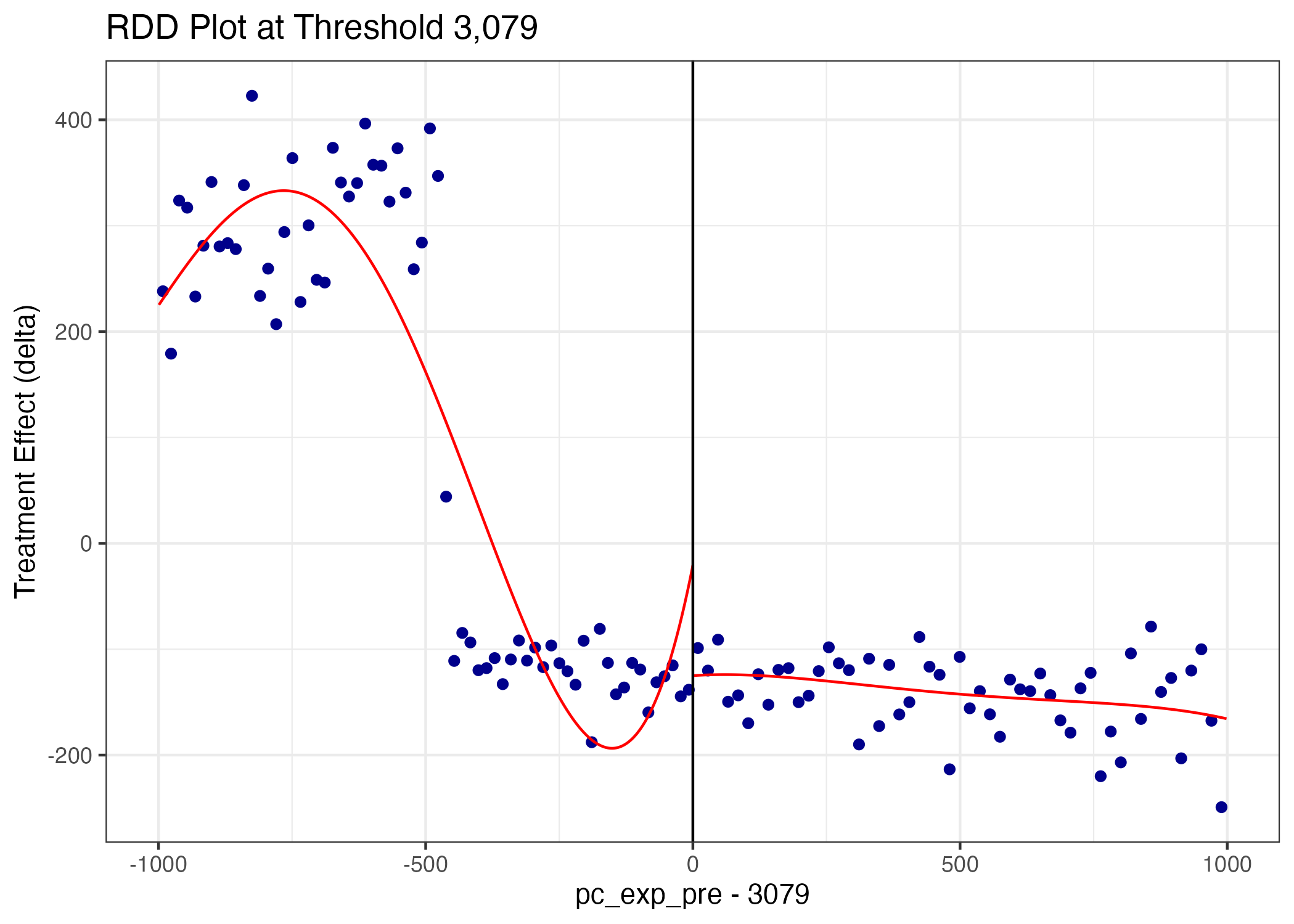}
\caption{RDD plot at the second estimated crossing point}\label{fig-RDD2}\end{subfigure}
\caption{RDD plots at the two estimated TEC crossing points (synthetic data).}\label{fig-RDD}
\end{figure}

The local discontinuity analysis at per capita expenditure equal to 1,658 (where the distributional effect turns from negative to positive) shows a positive and statistically significant jump (Figure~\ref{fig-RDD1}). The analysis at per capita expenditure of 3,079 (where the effect turns from positive to negative) yields a small positive jump that is not statistically significant, with a robust 95\% confidence interval of $[-18.5; 85.4]$ that includes zero (Figure~\ref{fig-RDD2}). The point estimates and conventional standard errors, together with the robust bias-corrected $p$-values and confidence intervals, are reported in Table \ref{tbl-rdd}. The significant jump at the first boundary supports the group-based pattern previously observed, whereas the insignificant estimate at the second shows that an estimated crossing need not coincide with a sharp discontinuity in individual responses.

[Table 4 approximately here.]

[Figure 3 approximately here.]

\section{A Real-Data Diagnostic: PROGRESA}\label{sec-progresa}
We close the empirical analysis with a short application to real-world data, where treatment effects are not known ex ante and distributional changes may be subtle. The purpose is diagnostic rather than a full demonstration. The synthetic section has already shown the full HDA and VDA workflow, while PROGRESA shows how the toolkit behaves when real data produce dominance in one comparison and weak, poorly localized sign changes in another. We use the public replication data of \citet{angelucci2009}, drawn from the experimental evaluation of PROGRESA, Mexico's flagship conditional cash transfer program. The evaluation design randomized program rollout across 506 rural villages (320 treatment, 186 control) in seven states, with eligibility within villages determined by a poverty index. Because randomization occurred at the village level, the unconfoundedness condition of Section \ref{sec-identification} holds by design, for both eligible households (the direct effect of the program) and ineligible households living in treatment villages (the indirect, or spillover, effect that is the focus of \citet{angelucci2009}). The outcome is monthly food consumption per adult equivalent, in pesos. We analyze the survey waves of November 1998 and November 1999. Heterogeneous impacts of PROGRESA have been documented before \citep{djebbari2008}. Our contribution here is to illustrate what the TEC toolkit adds when the relevant distributional effects are not large discontinuities but small changes in shape, sign, and timing.

[Figure 4 approximately here.]

[Table 5 approximately here.]

\begin{figure}[htbp]\centering
\includegraphics[width=\textwidth]{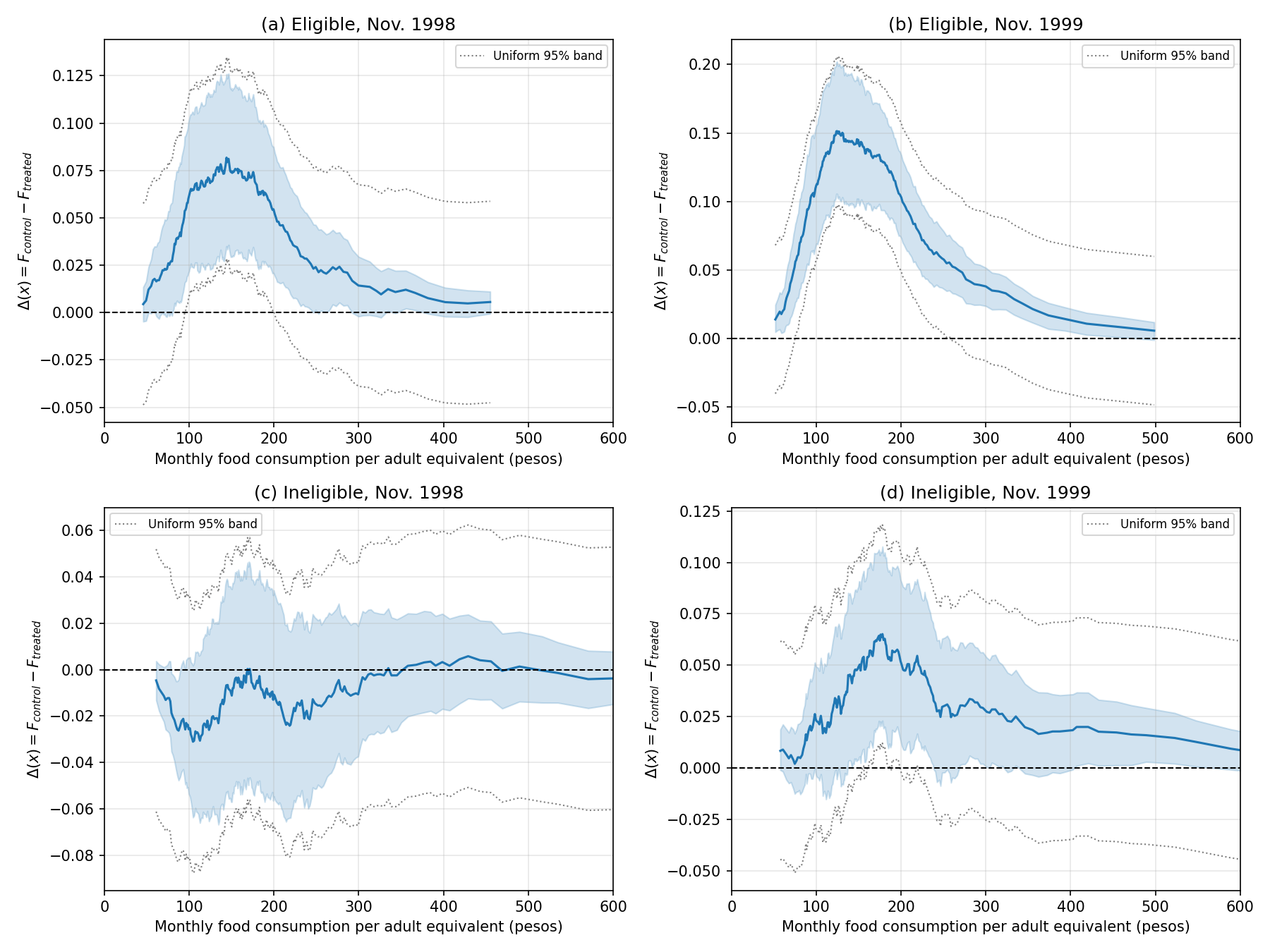}
\caption{Treatment Effects Curves for PROGRESA, monthly food consumption per adult equivalent, treatment versus control villages. Panels (a) and (b) eligible households, panels (c) and (d) ineligible households in the same villages. Shaded regions are pointwise 95\% confidence intervals from a village-level cluster bootstrap (999 replications), dotted lines are uniform (sup-norm) 95\% bands. Positive values indicate that treatment shifts the consumption distribution rightward.}\label{fig-progresa}
\end{figure}

The results, shown in Figure~\ref{fig-progresa} and Table \ref{tbl-progresa}, provide a compact real-data check on the decision rules developed above. For eligible households, the TEC is positive over the entire consumption distribution in both waves, and the crossing detector correctly finds no interior zeros (in the village-level cluster bootstrap, spurious crossings appear in 16\% and 1\% of replications, respectively). This is the dominance branch of the procedure. The estimated TEC is positive over the evaluated grid, consistent with first-order dominance, so program gains are positive across the distribution, though the village-level bootstrap shows this is less sharply identified in the tails. The TEC nevertheless adds information to the ATE of +15.8 pesos (1998) and +26.1 pesos (1999). The curves are hump-shaped, peaking at +0.08 and +0.15 around the 40th--50th percentile of consumption (roughly 125--145 pesos), and uniformly significant over 57\% and 82\% of the distribution, respectively. Even where the policy effect is positive throughout, the TEC shows that the gains are concentrated in the lower-middle of the distribution, much smaller in both tails, and stronger after longer exposure to the program.

The ineligible households display the subtler pattern for which the TEC is most useful. In November 1998, the mean spillover is slightly negative ($-6.2$ pesos) and the TEC hovers around zero. It is significantly negative (pointwise) over about 4\% of the grid, concentrated around the bottom decile, and the detector flags five interior crossings, scattered across the middle and the upper tail. A mean effect alone would summarize this as a small loss. A dominance test alone would emphasize the absence of a clean ordering. The TEC shows something more delicate: weak negative spillovers at the bottom, near-zero effects elsewhere, and sign changes that are visible but too flat to localize. At the first detected crossing (172 pesos, $\hat{p}_0 = 0.46$), the estimated slope is $-1.1 \times 10^{-4}$ and the non-tangentiality test cannot reject a flat crossing. Because PROGRESA randomized at the village level, we quantify the crossing's location with a village-level cluster bootstrap rather than with the analytic Section \ref{sec-inference} variances, which assume independent samples and are anti-conservative under clustering; the cluster bootstrap detects an interior crossing in 89\% of replications but places the first one anywhere in $[78;\, 431]$ pesos, an interval spanning much of the observed consumption range. The procedure thus detects a candidate crossing whose near-zero slope and very wide cluster-bootstrap interval leave its location essentially unidentified. The conclusion is not that there is a sharp threshold, but that short-run spillovers are weakly negative at the bottom of the distribution and that any sign switch is too gradual to support HDA or VDA at an estimated cutoff. By November 1999, the picture has changed. The spillover is positive on average (+26.3 pesos) and the TEC is positive everywhere, consistent with the insurance and credit channel documented by \citet{angelucci2009}, through which transfers reach ineligible households via gifts and loans.

The case of PROGRESA shows the diagnostic value of stopping when crossings are too flat to support threshold interpretation. When the estimated TEC does not cross and the bootstrap uncertainty is modest, the pattern is consistent with dominance over the evaluated range and shows where gains are concentrated--information invisible to the ATE. When crossings exist but are flat, the inference procedure reports that thresholds are not identified, protecting the researcher from spurious discontinuity analyses. Comparing TECs across waves also traces the dynamics of subtle distributional effects, here the transition of spillovers from weakly negative at the bottom to broadly positive as village credit and insurance networks absorbed the inflow of transfers.

\section{TEC Versus Stochastic Dominance and Quantile Regression Analyses}\label{tec-versus-sd-and-qr-analyses}
The advantage of studying TEC as opposed to ATEs is clear. Similarly to the motivation for conducting distributional treatment analyses like quantile regressions, the TEC provides a more nuanced picture of marginal distributional effects across the distribution of outcomes that is particularly relevant when these effects switch sign along the distribution.

The TEC analysis presented here also offers a complementary perspective to traditional stochastic dominance (SD) and quantile regression (QR) approaches for evaluating treatment effects. Stochastic dominance tests would result in no dominance across the entire distribution in our example, although they would be able to identify the crossing points of the two CDFs. However, as we showed, graphical illustrations would not clearly identify outcome regions with reversed marginal distributional effects, and the distribution of individual effects would not be identified without stronger assumptions. Standard SD tests \citep{davidson2000,barrett2003} provide p-values for dominance but do not directly quantify the magnitude or direction of differences at each point.

TEC analysis and quantile regression address distributional effects from different angles. Quantile regression estimates conditional quantile treatment effects (QTEs), answering questions like ``what is the effect of treatment on the 25th percentile of the outcome distribution, conditional on covariates?'' This approach is powerful for understanding how treatment effects vary across the outcome distribution while controlling for observable characteristics, and benefits from well-developed inference procedures including uniform confidence bands \citep{chernozhukov2018generic}. However, QR focuses on specific quantiles chosen by the researcher, requires specifying a model for covariate adjustment, and does not naturally highlight where effects change sign. TEC analysis, by contrast, provides a nonparametric visualization of the entire distribution of treatment effects without requiring covariate specification or quantile selection. The TEC directly reveals sign changes and their locations, making it particularly suited for identifying heterogeneous subgroups and motivating further analysis. While QR excels at estimating precise effect magnitudes at specific quantiles with covariate adjustment, TEC analysis excels at revealing the overall pattern of distributional effects and detecting discontinuities that warrant further investigation.

Causal forests \citep{wager2019,athey2019grf} excel at discovering which observable characteristics predict treatment effect heterogeneity, providing variable importance measures and enabling researchers to understand \emph{why} effects differ across individuals. However, causal forests estimate effects in covariate space rather than outcome space. They answer ``how does the treatment effect vary with age, education, or income?'' rather than ``how does the treatment effect vary across the outcome distribution?''. TEC analysis operates in outcome space, directly showing where in the distribution of outcomes the marginal distributional effect changes sign. The two approaches are thus complementary. TEC analysis can first identify that marginal distributional effects are positive in middle-income regions and negative in low- and high-income regions, while causal forests can then investigate which covariates explain membership in these regions and how CATEs vary within them. Indeed, our HDA framework explicitly combines these approaches, using TEC-identified groups as the basis for causal forest analysis to uncover the drivers of effect heterogeneity. More generally, these approaches are complementary rather than mutually exclusive. A comprehensive analysis of DTEs might use SD tests to establish whether clear dominance exists, TEC analysis to identify sign changes and define groups, QR to estimate covariate-adjusted effects at quantiles of interest, and causal forests within the HDA framework to explain the resulting heterogeneity.

\section{Conclusions}\label{conclusions}
This paper developed a toolkit for detecting and diagnosing distributional sign changes built on the TEC, first proposed by \citet{Verme2010TEC}. The TEC can be used to compare distributions of outcomes in impact evaluations when treatment effects are heterogeneous and distributional curves cross. It provides a simple graphical tool to highlight areas where marginal distributional effects switch sign along the distribution of outcomes. This information is useful for conducting horizontal and vertical discontinuity analyses that can complement distributional treatment effect evaluations based on quantile regressions.

The methodological contribution of this paper lies in formalizing the HDA and VDA frameworks as systematic approaches to investigating treatment effect heterogeneity, and in adapting crossing-point inference into two complementary inferential components: a crossing-location procedure for detecting and locating candidate crossings, and a Wald test for their non-tangentiality. HDA uses causal forests to compare groups located in regions with opposite marginal distributional effects, while VDA examines the neighborhoods around sign-change points. Both approaches can yield insights into the mechanisms driving heterogeneous treatment effects when the crossings are sufficiently well identified. The Monte Carlo evidence shows that the crossing-point estimator is approximately unbiased and $\sqrt{n}$-consistent, that spurious crossings are rare in moderate samples while flat crossings require large samples to be detected, and that the proposed confidence intervals are conservative for steep and well-separated crossings but can under-cover for flat crossings in small samples, with the explicit bias correction of the local slope restoring near-nominal calibration of the crossing-location interval.

Several limitations should be noted. First, the VDA approach relies on endogenously determined thresholds, which differs from standard RDD assumptions and requires careful interpretation. Second, the synthetic-data exercise, while illustrative, imposes treatment effects by construction, and the Monte Carlo study covers the crossing-point procedures but not the downstream HDA and VDA analyses. The PROGRESA diagnostic partially addresses this limitation by showing how the toolkit behaves on real experimental data. It indicates distributional dominance where it holds, but also diagnoses subtle spillover effects whose sign changes are too flat for threshold-based analyses to be warranted. This is an important practical lesson. In applied work, the most policy-relevant distributional changes may not appear as sharp discontinuities. They may instead be weak, localized, and dynamic, requiring tools that can reveal them without over-interpreting their thresholds. Future work should develop formal inference procedures that account for threshold estimation uncertainty in the VDA, and identify applications with steep, well-identified crossings where the full HDA and VDA machinery can be deployed on real data.

The TEC is easy to compute, and its uncertainty is straightforward to quantify with tools matched to the object of interest: bootstrap confidence bands for the curve itself, analytic normal-approximation intervals for the location and slope of each crossing, and a cluster bootstrap when treatment is randomized in clusters. We hope this toolkit proves useful for researchers and policymakers seeking to understand not just average treatment effects, but the full distribution of impacts and the factors that shape who benefits most, who benefits least, and where policy effects may reverse.

\section*{Acknowledgments}
The authors wish to thank colleagues at the Department of Statistical Science of the University of Bologna for useful informal discussions on the main ideas proposed by this paper. All remaining errors are ours. 

\section*{Reproducibility Statement}
All results, tables, and figures were produced with R and Python. The replication package lists the required packages and their versions and does not install or upgrade packages at run time, so that replication leaves the computational environment unchanged.

\section*{Data Availability Statement}
The synthetic data used in this paper are publicly available from the World Bank Microdata Library at \url{https://microdata.worldbank.org/index.php/catalog/5906}. The PROGRESA data are from the publicly available replication package of \citet{angelucci2009}, American Economic Association (\url{https://doi.org/10.3886/E113289V1}).

\section*{Conflict of Interest}
The authors declare no conflict of interest.

\newpage

\section*{Tables}

\begin{table}[htbp]\centering
\caption{Monte Carlo performance of the crossing-point estimator (1,000 replications; $n$ is the pooled sample size, with equal groups $n_A=n_B=n/2$). Detection is the share of replications in which the number of detected interior crossings equals the true number. For the placebo design it is the share of replications with at least one spurious interior crossing (false positives). These are, respectively, the detection probability and the false-positive rate of the crossing-detection rule. Bias, RMSE, SE ratio (mean estimated standard error over the Monte Carlo standard deviation), and coverage of nominal 95\% Wald intervals are computed conditional on correct detection.}
\label{tbl-mc}
\resizebox{\textwidth}{!}{%
\begin{tabular}{lrrrrrrrr}
\toprule
Design & $n$ & True $x_0$ & True $\Delta'(x_0)$ & Detection & Bias & RMSE & SE ratio & Coverage \\
\midrule
Single, steep & 1,000 & 0.670 & 0.183 & 94.1\% & 0.027 & 0.169 & 1.29 & 97.7\% \\
Single, steep & 4,000 & 0.670 & 0.183 & 99.8\% & 0.012 & 0.085 & 1.07 & 97.6\% \\
Single, steep & 16,000 & 0.670 & 0.183 & 100.0\% & 0.003 & 0.043 & 0.94 & 93.0\% \\
Single, flat & 1,000 & 0.607 & 0.053 & 22.1\% & 0.052 & 0.535 & 0.83 & 87.3\% \\
Single, flat & 4,000 & 0.607 & 0.053 & 47.3\% & 0.042 & 0.286 & 0.88 & 89.4\% \\
Single, flat & 16,000 & 0.607 & 0.053 & 86.2\% & 0.014 & 0.137 & 0.94 & 93.9\% \\
Double crossing (crossing 1) & 1,000 & 0.376 & -0.543 & 73.1\% & 0.009 & 0.045 & 8.71 & 100.0\% \\
Double crossing (crossing 2) & 1,000 & 0.967 & 0.326 & 73.1\% & -0.010 & 0.092 & 1.64 & 99.0\% \\
Double crossing (crossing 1) & 4,000 & 0.376 & -0.543 & 90.1\% & 0.002 & 0.022 & 3.08 & 100.0\% \\
Double crossing (crossing 2) & 4,000 & 0.967 & 0.326 & 90.1\% & -0.004 & 0.049 & 1.15 & 98.3\% \\
Double crossing (crossing 1) & 16,000 & 0.376 & -0.543 & 99.4\% & 0.001 & 0.011 & 2.08 & 100.0\% \\
Double crossing (crossing 2) & 16,000 & 0.967 & 0.326 & 99.4\% & -0.001 & 0.024 & 1.01 & 95.6\% \\
Placebo (no crossing) & 1,000 & -- & -- & 14.4\% (false pos.) & -- & -- & -- & -- \\
Placebo (no crossing) & 4,000 & -- & -- & 0.1\% (false pos.) & -- & -- & -- & -- \\
Placebo (no crossing) & 16,000 & -- & -- & 0.0\% (false pos.) & -- & -- & -- & -- \\
\bottomrule
\end{tabular}%
}
\end{table}

\begin{table}[htbp]\centering
\caption{Conventional versus bias-corrected slope estimator at the Silverman bandwidth (steep single-crossing design, $n=4{,}000$, 1,000 replications). Slope ratio is the mean estimated slope over the true slope. SE ratio is the mean estimated standard error over the Monte Carlo standard deviation. Coverage is for the nominal 95\% Wald interval for the crossing location.}
\label{tbl-mcbw}
\begin{tabular}{lrrr}
\toprule
Slope estimator & Slope ratio & SE ratio & Coverage \\
\midrule
Conventional & 0.67 & 1.50 & 99.7\% \\
Bias-corrected & 0.93 & 1.05 & 96.6\% \\
\bottomrule
\end{tabular}
\end{table}

\begin{table}[htbp]\centering
\caption{Crossing-point inference for the synthetic data. For each crossing the table reports the Crossing-location results (the estimated location $\hat{x}_0$, $\hat{p}_0$, its standard error, and 95\% confidence interval) and the C2 non-tangentiality test (the estimated local slope $\hat{\Delta}'(\hat{x}_0)$, its standard error, and the two-sided Wald $p$-value for $H_0:\Delta'(x_0)=0$).}
\label{tbl-crossings}
\resizebox{\textwidth}{!}{%
\begin{tabular}{lrrrrrrr}
\toprule
Comparison & Crossing $\hat{x}_0$ & $\hat{p}_0$ & SE($\hat{x}_0$) & 95\% CI & Slope $\hat{\Delta}'(\hat{x}_0)$ & SE($\hat{\Delta}'$) & $p$-value \\
\midrule
Pre vs post & 1,658 & 0.299 & 35 & [1,589; 1,726] & 1.87e-04 & 9.52e-06 & $<0.001$ \\
Pre vs post & 3,079 & 0.696 & 75 & [2,931; 3,227] & -8.69e-05 & 7.57e-06 & $<0.001$ \\
Control vs treated & 1,669 & 0.302 & 26 & [1,617; 1,720] & 3.54e-04 & 1.12e-05 & $<0.001$ \\
Control vs treated & 2,996 & 0.677 & 45 & [2,908; 3,083] & -2.11e-04 & 1.14e-05 & $<0.001$ \\
\bottomrule
\end{tabular}%
}
\end{table}

\begin{table}[htbp]\centering
\caption{Local linear, RDD-inspired estimates of the jump in the treatment effect at the estimated crossing points (synthetic data; triangular kernel; rdrobust bandwidth selection). The estimate and standard error are the conventional local-linear values; the $p$-value and confidence interval use the robust bias-corrected inference of Calonico, Cattaneo and Titiunik.}
\label{tbl-rdd}
\begin{tabular}{lrrrc}
\toprule
Threshold & Conv. estimate & Conv. SE & Robust $p$ & Robust 95\% CI \\
\midrule
1,658 & 254.5 & 17.8 & 0.0000 & [208.1; 290.3] \\
3,079 & 26.5 & 22.4 & 0.2068 & [-18.5; 85.4] \\
\bottomrule
\end{tabular}
\end{table}

\begin{table}[htbp]\centering
\caption{TEC summary statistics for PROGRESA. ATE is the raw difference in mean consumption (treated minus control, pesos per adult equivalent). Crossings are interior zeros of the TEC detected by the crossing-point procedure. Pointwise significance is the share of the evaluation grid where the TEC differs significantly from zero; uniform significance is the share where the TEC is significantly positive (uniform sup-norm band). Both are at the 95\% level.}
\label{tbl-progresa}
\resizebox{\textwidth}{!}{%
\begin{tabular}{lrrrrrr}
\toprule
Sample & $n_C$ & $n_T$ & ATE & Crossings & Pointwise sig. & Uniform sig. (positive) \\
\midrule
(a) Eligible, Nov. 1998 & 4,162 & 6,811 & +15.8 & 0 & 84\% & 57\% \\
(b) Eligible, Nov. 1999 & 4,047 & 6,510 & +26.1 & 0 & 100\% & 82\% \\
(c) Ineligible, Nov. 1998 & 1,801 & 2,842 & -6.2 & 5 & 4\% & 0\% \\
(d) Ineligible, Nov. 1999 & 1,711 & 2,575 & +26.3 & 0 & 56\% & 18\% \\
\bottomrule
\end{tabular}%
}
\vspace{2pt}
{\footnotesize \textit{Note:} Crossing-point inference uses a village-level cluster bootstrap (999 replications). The share of bootstrap replications with at least one interior crossing is 16\%, 1\%, 89\%, 55\% for panels (a)--(d); the 95\% cluster-bootstrap interval for the first ineligible-1998 crossing (panel c) is [78;\,431] pesos.}
\end{table}

\newpage

\section*{Figures}

\end{document}